\documentstyle{article}
\newcommand{\Be}{\begin{equation}}
\newcommand{\Ee}{\end{equation}}
\newcommand{\ybar}{\bar{\psi}}
\newcommand{\n}{\nu}
\newcommand{\nbar}{\bar\nu}
\newcommand{\y}{\psi}
\newcommand{\half}{{\scriptsize\frac{1}{2}}}
\newcommand{\dslash}{/\!\!\!\partial}
\newcommand{\Bea}{\begin{eqnarray}}
\newcommand{\Eea}{\end{eqnarray}}
\newcommand{\Eq}[1]{equation~(\ref{#1})}

\newcommand{\Ealf}{\nonumber\\}
\newcommand{\sgn}{\mbox{sgn}}
\begin{document}
\thispagestyle{empty}
\centerline{\normalsize\bf }
\baselineskip=22pt
\centerline{\normalsize\bf MSW-like 
enhancements without matter}
\baselineskip=13pt
\centerline{\footnotesize G. J. Stephenson Jr.,}
\baselineskip=13pt
\centerline{\footnotesize\it Department of Physics 
\& Astronomy,
 University of New Mexico}
\centerline{\footnotesize\it Albuquerque, New 
Mexico 87131}

\vspace*{0.3cm}
\centerline{\footnotesize and}
\vspace*{0.3cm}
\centerline{\footnotesize T. Goldman,}
\baselineskip=13pt
\centerline{\footnotesize\it  Theoretical Division, 
Los Alamos  
National
Laboratory}
\baselineskip=12pt
\centerline{\footnotesize\it Los Alamos, New 
Mexico 87545}
\vspace*{0.3cm}
\centerline{\footnotesize and}
\vspace*{0.3cm}
\centerline{\footnotesize B. H. J. McKellar}
\baselineskip=13pt
\centerline{\footnotesize\it School of Physics, 
University of Melbourne}
\baselineskip=12pt
\centerline{\footnotesize\it Parkville, Victoria 
3052, Australia}

\begin{abstract}

We study the effects of a scalar field,
coupled only to neutrinos, on oscillations
among weak interaction current eigenstates.
The effect of a real scalar field
appears as effective masses for the
neutrino mass eigenstates, the same for
$\nbar$ as for $\n$.  Under some conditions,
this can lead to a vanishing of $\delta m^2$,
giving rise to MSW-like effects.  We discuss
some examples and show that it is possible
to resolve the apparent discrepancy in spectra
required by r-process nucleosynthesis in the
mantles of supernovae and by Solar neutrino
solutions.

\end{abstract}

\pagebreak
\setcounter{page}{1}

	The possibility of neutrino oscillations, 
the change of the
weak interaction content of mass eigenstates 
during propagation,
is influenced by the presence of normal matter 
through the MSW
process~\cite{Wolf, M&S}.  This is important in 
the analysis 
of the Solar Neutrino problem~\cite{Bethe, 
Rosen} and in
the analysis of nucleosynthesis in 
supernovae~\cite{Fuller}.
These analyses would seem to require different 
assumptions
about the mass spectrum for the neutrinos.

We have recently examined the possibility that,
in addition to the 
Standard Model interactions, neutrinos interact 
with each
other through a weakly coupled, extremely light 
scalar
field~\cite{Clouds}.  In that study we showed 
that, for
a wide range of parameters consistent
with known phenomena,
it is possible that neutrinos form clouds in the 
early
Universe, that those clouds could influence the
evolution of structures on stellar scales, and that 
such
clouds could have observable consequences.  In 
this
Letter we wish to present the effects that such 
clouds
or other aggregates of neutrinos
could have on oscillation phenomena.  In 
particular, we
will discuss the possibility of enhanced 
oscillations in
regions with no normal matter, the interplay with 
the
usual MSW effect and the possibility that this
could resolve the apparent contradiction in 
required
spectra mentioned above.

	For the purpose of this Letter, we shall 
confine
ourselves the to case of one scalar field, although 
more
complicated scenarios are possible in principle.  
As
will be apparent, this already provides
for a great deal of complexity.  We are interested
in an effective theory for several (taken here as 
three)
neutrino mass eigenstates with vacuum masses
 $m_j$, where by vacuum mass we 
mean the physical mass which an isolated
neutrino would have, and a scalar boson $\phi$
with mass $m_s$.  At this point we shall ignore
 the weak interaction, although the effects of the
Standard Model are reflected in the vacuum 
masses.
The effective Lagrangian is
\Be 
{\cal L} = \sum_j\left[\ybar_j(i\dslash - m_j)\y_j
+g_j\ybar_j\y_j\phi\right]  
+\half\left[(\partial\phi)^2 -m_s^2\phi^2\right]
\Ee
which gives as the equations of motion
\Bea
\left[\partial^2 + m_s^2\right]\phi & = & 
\sum_jg_j\ybar_j\y_j \label{PHI}\\
\left[i\dslash - (m_j - g _j\phi ) \right]\y_j & = 
&0.\label{PSI}
\Eea

These equations are a special case of the equations 
of Quantum
Hadrodynamics~\cite{QHD} developed for the 
study of
relativistic nucleons interacting through the 
exchange of
scalar and vector mesons.  Following \cite{QHD}, 
as in
\cite{Clouds}, we invoke the Thomas-Fermi 
approximation
in which it is assumed that the fields locally take 
on the 
values of the infinite system appropriate to the 
neutrino
densities at that point.

One may view \Eq{PSI} as a definition of
an effective mass,
\Be
m^*_j = m_j -g_j\phi
\Ee
and it is useful to scale out the vacuum mass,
defining
\Be
y_j = m^*_j/m_j
\Ee
or
\Be
y_j = 1-\frac{g_j}{m_j}\phi\label{yheavy}
\Ee
With more than one mass eigenstate, it is possible 
for some $m_j^*$ to become negative.
The change of 
sign is no problem for a
fermion and can be removed by a redefinition
of phase (related to its properties under
charge conjugation~\cite{DBD}), we shall
retain the possibility of a negative sign as it makes 
the following discussion
easier to follow.  When any confusion would 
arise, we explicitly indicate signs and  absolute 
values, as appropriate.

If we now specialize to spherical distributions of 
neutrinos
and treat the neutrino distributions as zero 
temperature 
Fermi gases, \Eq{PHI} may be used to generate a 
radial
equation for the $\phi$ field
\Be
\frac{d^2\phi}{dr^2}+\frac{2}{r}\frac{d\phi}{dr}
+ \phi = \frac{1}{m^2_s}\sum_j g_j \ybar_j\y_j. 
\label{Dephi}
\Ee
The scalar density $\ybar_j\y_j$
is given by
\Bea
\ybar_j\y_j & = & \frac{4\pi w}{(2\pi)^3}
\int_0^{k_F} k^2dk\left|\frac{m^*_j}
{\sqrt{k^2+{m^*_j}^2}}\right| \Ealf
& = & \frac{4\pi 
w{m_j}^3\left|y_j\right|}{(2\pi)^3}
\left[x_{F_j}|e_{F_j}| -
y_j^2 \ln\frac{x_{F_j}+|e_{F_j}|}{|y_j|}\right]
\Eea
where $x_{F_j} = k_{F_j}/m_j$ is the Fermi
momentum of the distribution of $\n_j$ in units
of its vacuum mass, $e_{F_j} = 
\sgn(y_i)\sqrt{\left(
x_{F_j}^2 + y_j^2\right)}$ and $w$ is the number
of neutrino states contributing to the scalar 
density.
For Majorana neutrinos $w = 2$  and $w = 4$ for
Dirac neutrinos.
As usual, the number density for each neutrino 
state, 
$\rho_j$, is related to the 
Fermi momentum $k_{F_j}$ by
\Bea
\rho_j &=& \frac{1}{6\pi^2}{k_{F_j}}^3\\
&=& \frac{m_j^3}{6\pi^2}x_{F_j}^3
\Eea

It is convenient to use \Eq{yheavy} to recast
\Eq{Dephi} as an equation for one of the $y_j$
and to rescale the radial variable as $z = m_sr$
~\cite{Clouds}.  We choose to follow the 
scaled effective
mass of the heaviest neutrino, $y_h$.  Defining the
ratios
\Be
r_j = m_j/m_h
\Ee
and the parameter
\Be
K_0=\frac{{g_h}^2wm_h^2}{2\pi^2m_s^2}
\Ee
we obtain
\begin{equation}
\frac{d^2y_h}{dz^2} + 
\frac{2}{z}\frac{dy_h}{dz} 
= -1+y_h( 1+\frac{K_0}{2}F)
\end{equation}
where
\begin{equation}
F=\sum_j\left|\frac{y_j}{y_h}\right|\frac{g_j}{g_h
}
\left[|e_{F_j}| x_{F_j} 
-y_j^2\ln\frac{|e_{F_j}|+x_{F_j}}{|y_j|}\right]  
\end{equation}
In these equations, the $|e_{F_j}|$ serve as 
Lagrange multipliers to fix the total number, 
$N_j$,  of
each mass eigenstate and are constant throughout 
the distribution~\cite{QHD}.

The details of any distribution about some center 
depend
on many things, including $N_j, m_j $ and $ g_j$.  
For example, as discussed
in~\cite{Clouds}, if the couplings were to be
proportional to the vacuum masses, all of
the $y_j$ would be the same at every point in
the distribution and all of the effective masses
would be positive.  Since the vacuum masses 
depend on other interactions as well as the 
scalar field, this seems an unlikely scenario.

The richness of the system can be demonstrated 
with another simple model, one in which the
various couplings are all equal to the same
constant $g$.  Beginning for simplicity with two 
mass eigenstates, let the vacuum mass of the 
heavier be
denoted by $m_h$ and that of the lighter by
$m_l$ with a ratio given by
\Be
r = \frac{m_l}{m_h},
\Ee
  In this case, the
shift from the vacuum mass to the effective 
mass is the same for both neutrinos,
\Bea
m^*_h &=& m_h - g\phi\\
m^*_l &=& m_l - g\phi\\
\Delta y &=& g\phi/m_h\\
y_h &=& 1-\Delta y\\
y_l &=& 1 - \Delta y/r\\
&=& 1 - (1-y_h)/r
\Eea
If,
at the center of the distribution, the shift
is large enough, the effective mass of the
lighter neutrino changes sign and can have
a magnitude greater than its vacuum mass.  In that 
case,
 it is energetically
favorable for the light neutrino to move
to a region with smaller scalar field, 
so the center of the distribution
will have only the heavier neutrino present. As 
long as $y_l < -|e_{F_l}|$, the density of
light neutrinos will be zero.  The density of 
heavy neutrinos goes to zero when $y_h > 
e_{F_h}$
so, if $1-(1-e_{F_h})/r < -|e_{F_l}|$ there will
be an annulus with no neutrinos in which $\phi$
decreases and $y_l$ increases.  When 
$y_l = -|e_{F_l}|$, the density of light
neutrinos becomes non-zero, and remains
so until $y_l > +|e_{F_l}|$.  Beyond this
radius there are no neutrinos and $\phi$ falls
to zero as $\exp(-z)/z$.

The importance of this for MSW-like effects is the
following.  Enhanced transitions between
mass eigenstates occur when the
propagators are equal for the same momentum,
i.e. when
\Be
{m^*_h}^2 - {m^*_l}^2 = 0
\Ee
In this example the equality is achieved when
\Be
m^*_l = -m^*_h,
\Ee
which occurs when
\Be
y_h = y^0_{lh} = (1-r)/2 \label{Cross}
\Ee
This condition can be achieved
in the region
with no neutrinos.  As an example of this,
we have revisited the case 
presented as Figure 9 of~\cite{Clouds},
studying the value of $\left[{m^*_h}^2
-{m^*_l}^2\right]/m_h^2$.  The result, for 
$r=0.1$, $e_{F_h} = 0.43$ and $|e_{F_l}| = 
0.9805$, is shown in Figure 1, in which we display 
the latter difference as well as the densities of the 
heavy and the light neutrinos.  The density scales 
are arbitrary and different for ease of viewing.  As 
indicated by \Eq{Cross}, in this
case the vanishing of the mass-squared
difference not only occurs without ordinary 
matter but at a location where the
density of neutrinos is zero.

The extension to three mass eigenstates is 
quite straightforward. Denote the three mass
eigenstates by $m_h$, $m_m$ and $m_l$ in an
obvious notation.  Define
\Bea
r_j& = &m_j/m_h\\
y_l&=&1-(1-y_h)/r_l\\
y_m &=& 1-(1-y_h)/r_m.
\Eea
Now it is possible, in principle, to have three
distinct 
effective level crossings, where $y_h$ takes the 
values
\Bea
y^0_{hl}& =& (1-r_l)/2\\
y^0_{hm}&=& (1-r_m)/2\\
y^0_{ml} &= &1-(r_l+r_m)/2.
\Eea

Note that this particularly simple pattern
arises because of the assumption that all
three mass states couple to the scalar field 
with the same strength.  In general, with
\Bea
\gamma_j & = & g_j/g_h,\\
y_j & = & 1-\frac{\gamma_j}{r_j}(1-y_h).
\Eea
The level crossings now occur for values
of  $y_h$ given by
\Bea
y^0_{hj} &= &\frac{\left(\gamma_j - r_j\right)}
{\left(1+\gamma_j\right)}, \quad\quad 
\mbox{and}\\
y^0_{ml} &=& 1 - \frac{\left(r_m + r_l\right)}
{\left(\gamma_m + \gamma_l\right)}.
\Eea
Since $y_h > 0$, there will be no crossing
of  $\n_j$ with $\n_h$ if $\gamma_j < r_j$
and no crossing of $\n_l$ with $\n_m$ if
$\left(\gamma_m + \gamma_l\right) <
\left(r_m + r_l\right)$.  In particular, for the
case where the couplings are proportional
to the vacuum masses there are no crossings
for finite density.

Since this change in the  values of the
effective masses is due to the presence of
a scalar field, it is the same for $\bar\n$
as for $\n$.  This is different from the
MSW effect which, being an energy
shift arising from a vector
exchange, has the opposite sign for $\n$
and $\bar\n$.  In the presence of both a
large density of neutrinos and a large
electron density, these two effects will
both come into play, allowing the crossings
for $\n$ and $\nbar$ to occur at different
radii.  

With this remark, one may understand the
source of
apparent conflict between the MSW effect on
Solar neutrinos and the requirements of r-process
nucleosynthesis in supernovae~\cite{Fuller}.
Assume, for this argument, three points.  First,
the usual prejudice about the spectrum holds.
This means that the lightest mass eigenstate has
the largest overlap with the electron neutrino.
Second, that the density of neutrinos (strictly, 
the strength of the scalar field) in the Sun is small
enough that the usual MSW effect is not severely
perturbed.  Third, that the density of neutrinos in
a supernova is sufficiently large that 
$y_l < -y_h$.  With these assumptions, the
radius in the supernova
 at which the neutrinos are degenerate
is smaller than the radius at which 
the anti-neutrinos are degenerate.  This means
~\cite{Fuller} that the light neutrinos are
reheated at a smaller radius, hence traverse
more matter and are more efficiently
recooled, than the light anti-neutrinos.
This can provide the required feature that
the effective temperature of $\nbar$ is
higher than that of $\n$
at the site of the r-process.

We should emphasize that this effect in
supernovae could occur, given a scalar
field coupled to neutrinos, even if the
coupling were too weak or the range too 
short to provide stable clouds~\cite{Clouds}.
The extremely large neutrino densities 
associated with supernovae would still be
able to alter the effective masses and 
could provide for the degeneracy in 
${m^*}^2$ discussed above.

Note that, even if there is no scalar field, 
the vector interaction with matter could be
large enough to produce
degeneracy for the anti-neutrinos.
In this case, however, the anti-neutrino 
degeneracy occurs at a smaller radius than
the neutrino degeneracy and the desired
shift in temperatures is not achieved.

Another phenomenon in which a neutrino density 
will produce a non-linear MSW-like effect 
through the standard model has been discussed in 
the literature~\cite{TMKS}.  In this case a 
neutrino interacts with the background neutrinos 
by $Z^0$ exchange.  While this effect can be 
incorporated into the QHD equations, we have not 
done so because at the neutrino densities in stars 
and supernovae, and for the range of parameters
we have considered, it is much smaller than
the scalar field effect we have discussed above.

In summary, we have shown that the
effect of a scalar
field coupled to neutrinos, expressed as
an effective mass, could lead to a 
degeneracy in the propagators of 
different mass eigenstates.  In objects
with a radially varying neutrino density,
be they supernovae, clouds of relic
neutrinos or other astronomical structures,
these degeneracies  would occur at particular
radii leading to enhanced transitions between
weak interaction eigenstates in the manner of
the MSW effect~\cite{Wolf,M&S,Bethe,Rosen}.
Unlike MSW, this effect is the same for $\n$
and $\nbar$, so the interplay between the
two, depending on both the matter density
and the neutrino density in a given region of
space, can produce effects which would 
appear, with MSW alone, to require different
mass spectra~\cite{Fuller}.

This work has been supported in part by the
United States National Science Foundation, 
the United States Department of Energy,
the Australian Research Council and the
Australian DIST.

\newpage
\section*{Figure Captions}
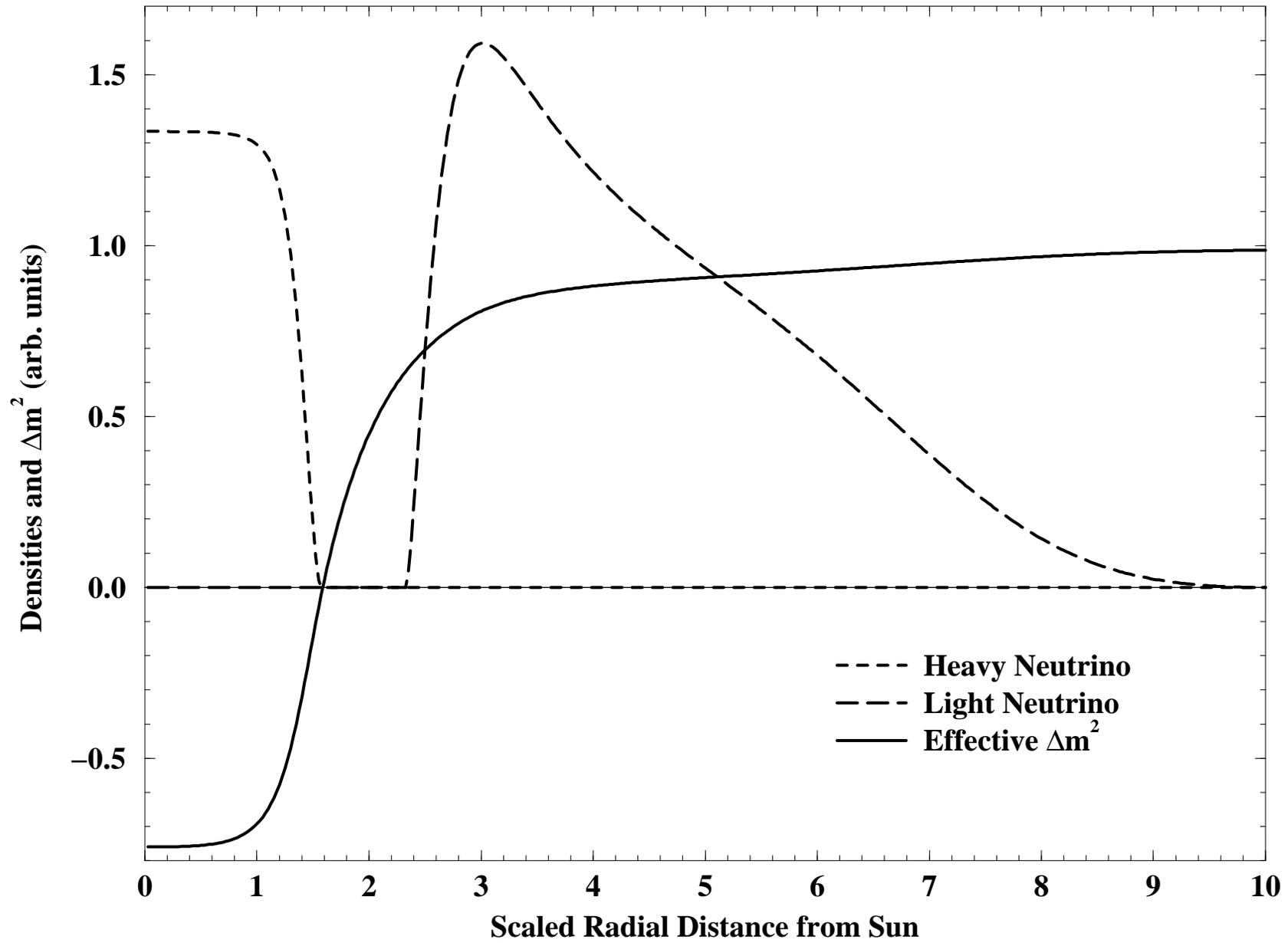
\noindent Figure 1.     Densities and  $\Delta 
{m^*}^2 = \left[{m^*_h}^2
-{m^*_l}^2\right]/m_h^2$  for two
neutrino mass eigenstates with constant coupling 
to the scalar  
field.
$K_0~=~400$ for the heavy neutrino and
$m_l/m_h~=~0.1$.  The 
long-dashed line is proportional to  the density for 
the  
heavy
neutrino, the dot-dashed line is proportional to  
the density of the light  
neutrino
and the solid line is proportional to the value of 
$\Delta {m^*}^2$.   The scaled Fermi energies are  
$e_{F_h} = 0.43$ and $|e_{F_l}| = 0.9805$,  and 
the numbers of the two neutrino mass eigenstates 
are approximately equal.  Although it is not 
apparent from the figure the point at which $\Delta 
{m^*}^2 = 0 $ lies outside the region in which the 
heavy neutrinos are found.

\end{document}